\documentclass[
reprint,
superscriptaddress,
amsmath,amssymb,
aps,
]{revtex4-2}

\usepackage{graphicx}
\usepackage{dcolumn}
\usepackage{bm}
\usepackage{hyperref}
\usepackage{float}

\begin{document}

\title{
Unified origin of negative energetic elasticity in a lattice polymer chain:\\
soft self-repulsion and bending stiffness
}

\author{Nobu C. Shirai}
\email{shirai@cc.mie-u.ac.jp}
\affiliation{
Center for Information Technologies and Networks, Mie University, Tsu, Mie 514-8507, Japan
}

\date{\today}

\begin{abstract}
We study a single lattice polymer chain under a fixed end-to-end distance, incorporating both Domb--Joyce (DJ) soft-core self-repulsion between polymer segments and a local bending-energy cost. 
By decomposing the stiffness into energetic and entropic contributions, we survey the parameter space defined by the self-repulsion strength and bending-energy cost. We find that the energetic contribution to stiffness is negative across the entire explored range, whereas the entropic contribution remains positive. 
These results unify two previously independent mechanisms of negative energetic elasticity---solvent-induced self-repulsion and bending stiffness---and demonstrate that either mechanism alone, as well as their combination, produces the same sign. 
Beyond this sign-level unification, we analyze the internal-energy scaling and show that, in the absence of the bending-energy term, the DJ (self-repulsion) limit exhibits a robust $(n-r)^{7/4}/n$ scaling collapse. In contrast, the introduction of finite bending stiffness progressively disrupts this scaling, providing an internal-energy-based diagnostic to distinguish between contributions from self-repulsion and bending stiffness.
\end{abstract}

\maketitle

\section{Introduction}
\label{sec:intro}
Negative energetic elasticity~\cite{YoshikawaSakai2021,FujiyabuSakumichi2021}, defined as the negative contribution of internal energy to the elastic stiffness, has been demonstrated to be robust across a wide range of gel systems~\cite{TangChen2023,AoyamaUrayama2023,LiuTang2024} and theoretical models~\cite{ShiraiSakumichi2023,DuarteRizzi2023,IwakiOzaki2024,HagitaSakumichi2023,ShiraiSakumichi2025}. 
Our previous analyses of the interacting self-avoiding walk and the Domb--Joyce (DJ) model established that solvent-induced self-repulsion between polymer segments alone is sufficient to generate this effect~\cite{ShiraiSakumichi2023,ShiraiSakumichi2025}. 
Independently, an exactly solvable random-walk model with a bending-energy cost, mapped onto an Ising chain, has shown that bending stiffness likewise gives rise to a negative energetic contribution to elasticity~\cite{IwakiOzaki2024}.
Although both classes of models predict the same negative sign, the differences between the solvent-induced and bending-induced mechanisms have not been systematically compared. 

To enable a systematic comparison of these mechanisms and to develop diagnostic criteria, we investigate the elastic response of a single lattice polymer chain at a fixed end-to-end distance in the presence of both self-repulsion and a local bending-energy cost. 
In the bending-only limit, the model admits an exact mapping onto a one-dimensional Potts chain (and, under axial restriction, onto an Ising chain), which may be regarded as a three-dimensional extension of the Ising-chain formulation introduced in Ref.~\cite{IwakiOzaki2024}.
Building on the DJ model~\cite{DombJoyce1972} augmented by a bending-energy term, we decompose the mechanical stiffness into energetic and entropic contributions and demonstrate that, over the explored parameter range, the energetic component is negative while the entropic component remains positive.
In particular, we exploit the scaling of the internal-energy with the slack $(n-r)$~\cite{ShiraiSakumichi2025} to construct a quantitative diagnostic: we show that, in the DJ limit, the internal energy collapses onto a robust master curve proportional to $(n-r)^{7/4}/n$; by contrast, increasing bending stiffness progressively breaks this scaling.

\section{Model and Method}
\label{sec:model}
\begin{figure}[t]
\centering
\includegraphics[width=0.99\linewidth]{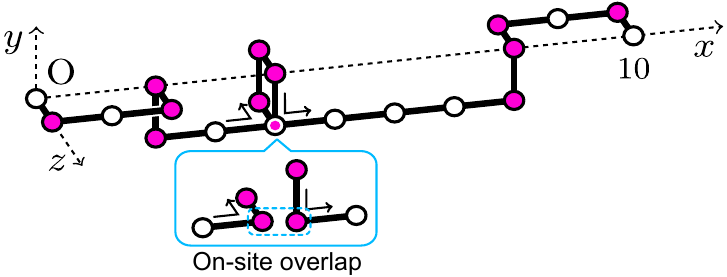}
\caption{
A 20-step random-walk configuration on a cubic lattice with endpoints fixed at the origin $\mathrm{O}$ and $(10,0,0)$ (on-axis constraint).
Pink-filled circles indicate chain segments contributing to the bending-energy cost.
The inset highlights an on-site overlap ($v_x = 2$), corresponding to two segments occupying the same lattice site, which contributes to the DJ overlap count $m=1$.
Local changes in the chain direction contribute to the bend count $b=13$.
}
\label{fig:config}
\end{figure}

An $n$-step random walk on a simple cubic lattice is defined as a sequence $\omega=[\omega(0),\ldots,\omega(n)]$ with $\omega(i)\in\mathbb{Z}^3$ and $|\omega(i+1)-\omega(i)|=1$. 
The lattice spacing is set to unity.
We fix the endpoints at $\omega(0)=(0,0,0)$ and $\omega(n)=(r,0,0)$.
The step vectors are defined as $\mathbf{r}_i\equiv \omega(i+1)-\omega(i)\in\{\pm \hat{\mathbf{x}},\pm \hat{\mathbf{y}},\pm \hat{\mathbf{z}}\}$ for $i=0,\ldots,n-1$, where $\hat{\mathbf{x}}$, $\hat{\mathbf{y}}$, and $\hat{\mathbf{z}}$ denote unit vectors along the coordinate axes.
We define the total energy as the sum of a DJ soft-core interaction term and a bending-energy term:
\begin{equation}
E(\omega) = \varepsilon\, m(\omega) +  \varepsilon_b\, b(\omega), 
\label{eq:Etot}
\end{equation}
where $\varepsilon\ge 0$ and $\varepsilon_b\ge 0$ (excluding the trivial case $\varepsilon=\varepsilon_b=0$), $m(\omega)$ denotes the DJ overlap count, and $b(\omega)$ denotes the bend count.
For each lattice site $x\in\mathbb{Z}^3$, let $v_x(\omega)$ denote the number of segments of $\omega$ occupying site $x$.
Then
\begin{equation}
m(\omega)=\sum_{x\in\mathbb{Z}^3}\binom{v_x(\omega)}{2}
=\sum_{x\in\mathbb{Z}^3}\frac{v_x(\omega)\,[v_x(\omega)-1]}{2},
\end{equation}
and
\begin{equation}
b(\omega)=\sum_{i=0}^{n-2}\left[1-\delta_{\mathbf{r}_i,\mathbf{r}_{i+1}}\right].
\end{equation}
Note that $0\le b\le n-1$.
In particular, $b=0$ occurs only for a perfectly straight configuration, which is possible only when $n=r$.
A representative configuration with $(n,r)=(20,10)$ is shown in Fig.~\ref{fig:config}.
The inset illustrates an on-site overlap ($v_x=2$, contributing $m=1$), and the depicted configuration has a total bend count of $b=13$.

Setting $\varepsilon=0$ in Eq.~\eqref{eq:Etot} leaves only the bending-energy term
\begin{equation}
\varepsilon_b\, b(\omega)=\varepsilon_b \sum_{i=0}^{n-2}\!\left[1-\delta_{\mathbf{r}_i,\mathbf{r}_{i+1}}\right]
= \varepsilon_b(n-1)-\varepsilon_b \sum_{i=0}^{n-2}\delta_{\mathbf{r}_i,\mathbf{r}_{i+1}},
\label{eq:Eb-potts}
\end{equation}
where each step vector $\mathbf{r}_i\in\{\pm \hat{\mathbf{x}},\pm \hat{\mathbf{y}},\pm \hat{\mathbf{z}}\}$ denotes one of the six lattice directions.
Up to an additive constant $\varepsilon_b(n\!-\!1)$, Eq.~\eqref{eq:Eb-potts} is equivalent to a one-dimensional ferromagnetic Potts chain defined along the walk, with $q=6$ states and coupling $J=\varepsilon_b$.
If the steps are restricted to a single axis (the $q=2$ case), we take $\mathbf{r}_i\in\{\pm\hat{\mathbf{x}}\}$ and define $\sigma_i\equiv \mathbf{r}_i\cdot\hat{\mathbf{x}}\in\{\pm1\}$.
In this case, $\delta_{\mathbf{r}_i,\mathbf{r}_{i+1}}=(1+\sigma_i\sigma_{i+1})/2$, and
\begin{equation*}
\varepsilon_b\, b(\omega)=\frac{\varepsilon_b}{2}\sum_{i=0}^{n-2}\!\left[1-\sigma_i\sigma_{i+1}\right]+\mathrm{const.},
\end{equation*}
which is exactly the Ising-chain Hamiltonian employed in Ref.~\cite{IwakiOzaki2024}. 
Under a fixed end-to-end vector, this bending-only limit corresponds to a 3D extension of the exactly solvable 1D model of Ref.~\cite{IwakiOzaki2024}.

At inverse temperature $\beta\equiv 1/(k_\mathrm{B}T)$, the partition function for fixed $n$ and $r$ is
\begin{equation}
Z=\sum_{m=0}^{\infty}\sum_{b=0}^{n-1} W_{n,m,b}(r)\,e^{-\beta(\varepsilon m + \varepsilon_b b)},
\end{equation}
where $W_{n,m,b}(r)$ denotes the number of $n$-step walks with end-to-end distance $r$, DJ overlap count $m$, and bend count $b$, and we define $W_{n,m,b}(r)=0$ whenever no such walk exists. 
For fixed $(n,r)$, $m$ is bounded above (an explicit tight upper bound can be constructed; see the Appendix on upper bounds in Ref.~\cite{ShiraiSakumichi2025}), so the sum over $m$ is effectively finite.

The free energy, internal energy, and entropy are given, respectively, by
\begin{eqnarray}
A&=&-(1/\beta)\ln Z,\\
U&=&\frac{1}{Z}\sum_{m=0}^{\infty}\sum_{b=0}^{n-1}(\varepsilon m+\varepsilon_b b)\,W_{n,m,b}(r)\,e^{-\beta(\varepsilon m+\varepsilon_b b)},\\
S&=&k_\mathrm{B}\beta(U-A).
\end{eqnarray}

We define the stiffness of the lattice polymer chain under the on-axis constraint.
In the continuum limit, the stiffness is given by the second derivative of the free energy: 
\begin{equation*}
\frac{\partial^2 A}{\partial r^2}=\frac{1}{\beta}\left[ \left( \frac{\frac{\partial Z}{\partial r}}{Z} \right)^2 
- \frac{\frac{\partial^2 Z}{\partial r^2}}{Z} \right].
\end{equation*}
Accordingly, we define the stiffness in finite-difference form as
\begin{eqnarray}
k &\equiv& \frac{1}{\beta}\left\{ \left[
\frac{1}{Z}\sum_{m=0}^{\infty}\sum_{b=0}^{n-1}
\frac{\varDelta W_{n,m,b}(r)}{\varDelta r}\, e^{-\beta(\varepsilon m + \varepsilon_b b)}
\right]^2 \right. \nonumber\\
&& \left.
- \frac{1}{Z}\sum_{m=0}^{\infty}\sum_{b=0}^{n-1}
\frac{\varDelta^2 W_{n,m,b}(r)}{\varDelta r^2}\, e^{-\beta(\varepsilon m + \varepsilon_b b)}
\right\},
\label{eq:k}
\end{eqnarray}
where the first- and second-order differences of $W_{n,m,b}(r)$ are given, respectively, by
\begin{equation*}
\varDelta W_{n,m,b}(r) \equiv \left[W_{n,m,b}(r+\varDelta r)- W_{n,m,b}(r-\varDelta r)\right]/2,
\end{equation*}
and
\begin{equation*}
\varDelta^2 W_{n,m,b}(r) \equiv W_{n,m,b}(r+\varDelta r) - 2W_{n,m,b}(r)+ W_{n,m,b}(r-\varDelta r).
\end{equation*}
Here, $\varDelta r = 2$ because admissible $r$ shares the parity of $n$.

We decompose the mechanical stiffness into its underlying physical contributions as $k=k_U+k_S$, where $k_U\equiv \partial^2 U/\partial r^2$ represents the energetic contribution and $k_S\equiv -T\,\partial^2 S/\partial r^2$ represents the entropic contribution. 
Using the thermodynamic relation $S=-\partial A/\partial T$, these contributions can be expressed as
\begin{eqnarray*}
k_S &=& T\frac{\partial k}{\partial T}= -\beta\frac{\partial k}{\partial \beta},\\
k_U &=& k-k_S.
\end{eqnarray*}
Both contributions are calculated using Eq.~\eqref{eq:k}.

We enumerate all $n$-step cubic-lattice walks with fixed endpoints $\omega(0)=(0,0,0)$ and $\omega(n)=(r,0,0)$, recording, for each trajectory $\omega$, the DJ overlap count $m(\omega)$ and the bend count $b(\omega)$.
The enumeration is performed via a depth-first recursion that appends one of the six lattice directions at each step and updates the site-occupancy map $v_x$ along the walk, following the approach of Ref.~\cite{ShiraiSakumichi2025}. 
Once a walk of length $n$ reaches $(r,0,0)$, the DJ overlap and bend counts are computed as follows:
$m=\sum_x v_x(v_x-1)/2$ and $b=\sum_{i=0}^{n-2}[1-\delta_{\mathbf{r}_i,\mathbf{r}_{i+1}}]$.

\section{Results}
\label{sec:results}
We first characterize the behavior of the energetic and entropic contributions to the stiffness across the parameter space, then present an intuitive interpretation of the bending contribution, and finally analyze the scaling behavior of the internal energy as a probe of the underlying microscopic mechanism.

\begin{figure}[t]
\centering
\includegraphics[width=0.99\linewidth]{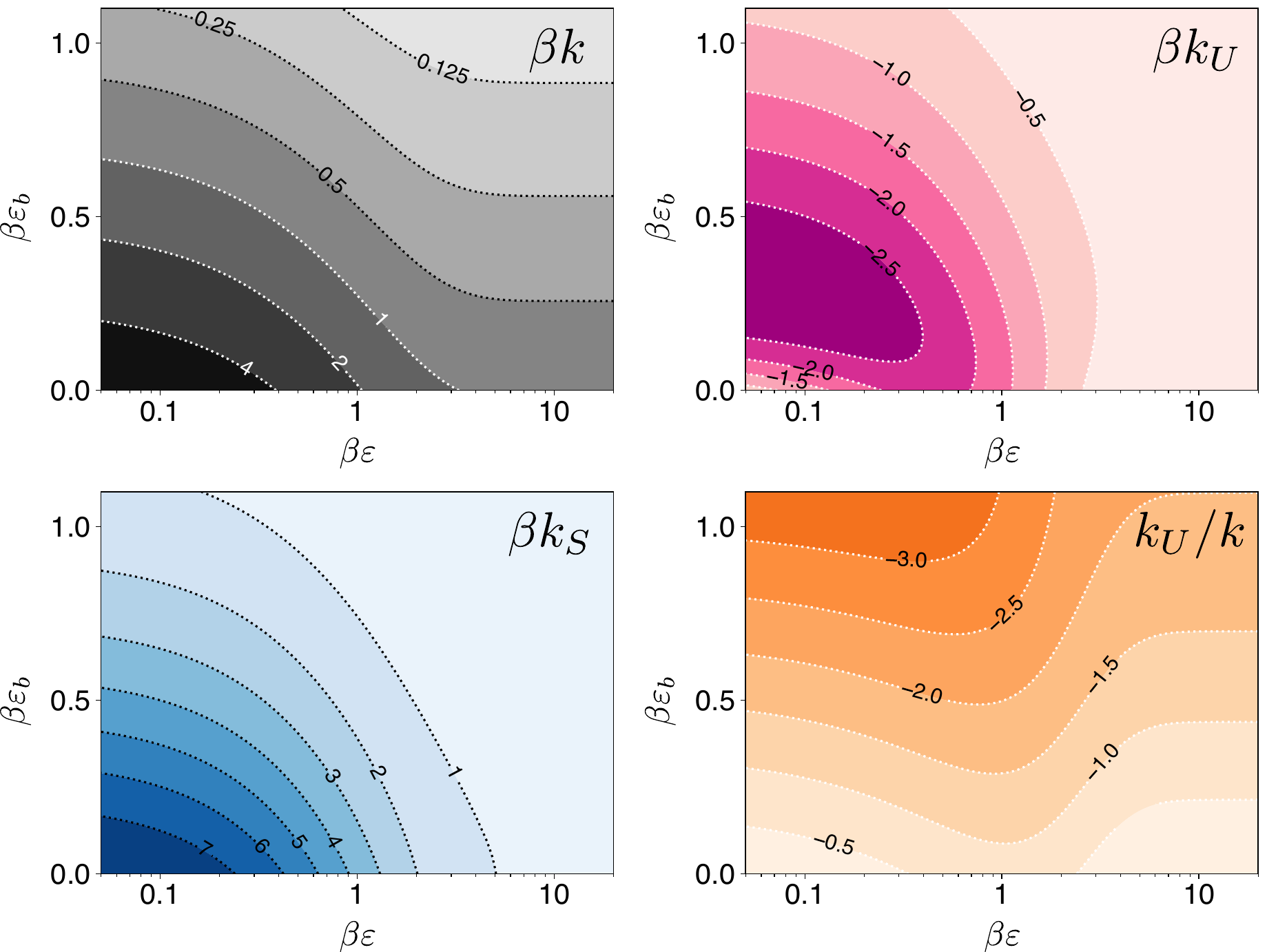}
\caption{
Contour plots of the stiffness decomposition in the model: $\beta k$, $\beta k_U$, $\beta k_S$, and $k_U/k$ across the $(\beta\varepsilon, \beta\varepsilon_b)$ plane for a chain with $(n,r)=(20,10)$. 
The $x$-axis is displayed on a logarithmic scale. 
Dotted contours represent lines of constant value in all panels. 
Negative energetic elasticity is observed throughout the domain, where either $\varepsilon > 0$ or $\varepsilon_b > 0$, as evident in both the $\beta k_U$ and $k_U/k$ panels.
}
\label{fig:landscape}
\end{figure}

Figure~\ref{fig:landscape} presents the decomposition of the stiffness across the $(\beta\varepsilon,\beta\varepsilon_b)$ plane for a representative chain with $(n,r)=(20,10)$. 
Over the entire explored parameter range, the energetic contribution $k_U$ is negative, whereas the entropic contribution $k_S$ remains positive. 
The total stiffness $k$ is determined by the competition between these two contributions.
The ratio panel $k_U/k$ confirms that the energetic fraction remains negative throughout the plane, indicating that either self-repulsion or bending stiffness alone is sufficient to produce negative energetic elasticity, and that their combination does not reverse this sign.

\begin{figure}[t]
\centering
\includegraphics[width=0.99\linewidth]{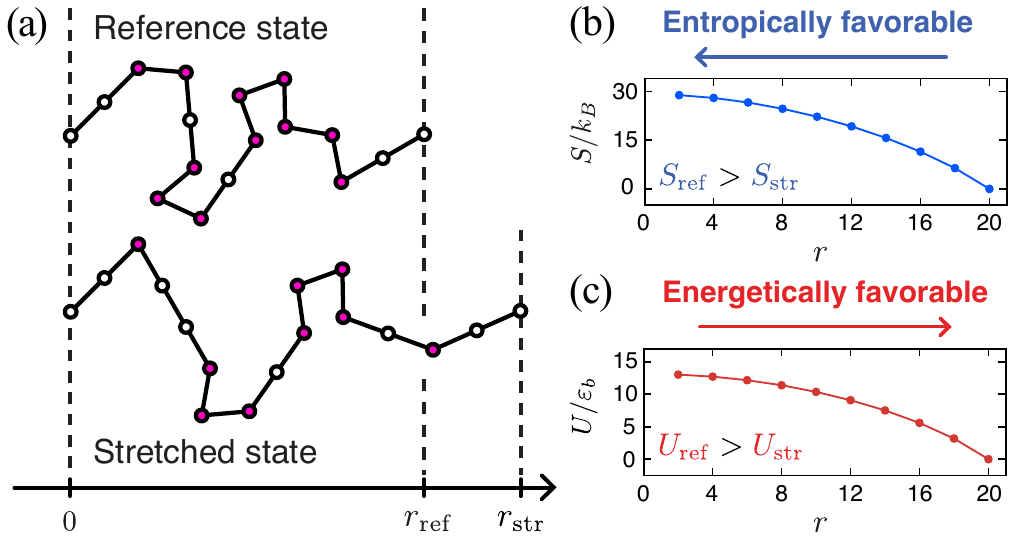}
\caption{
Illustration of negative energetic elasticity originating from the bending-energy cost ($\varepsilon_b>0$ and $\varepsilon=0$).
(a) Schematic comparison of chain conformations at two imposed end-to-end distances: $r_\mathrm{ref}$ for the reference state (top) and $r_\mathrm{str}\,(>r_\mathrm{ref})$ for the stretched state (bottom).
Pink-filled circles indicate local bends, i.e., segments contributing to the bending-energy cost.
At the larger imposed distance $r_\mathrm{str}$, typical conformations contain fewer bends, so the average bending contribution is reduced relative to $r_\mathrm{ref}$.
(b) Entropy $S/k_B$ as a function of $r$, showing that smaller extensions are more entropically favorable ($S_\mathrm{ref}>S_\mathrm{str}$).
(c) Internal energy $U/\varepsilon_b$ as a function of $r$, showing that larger extensions are energetically favorable ($U_\mathrm{ref}>U_\mathrm{str}$).
Parameters are set to $\varepsilon=0$, $\beta\varepsilon_b=1$, and $n=20$.
}
\label{fig:diagram}
\end{figure}

Figure~4 of Ref.~\cite{ShiraiSakumichi2025} provides a schematic illustration of how negative energetic elasticity arises from self-repulsion in lattice polymers. 
In the present model, Figure~\ref{fig:diagram} serves an analogous role for the bending contribution, focusing on the bending-only limit ($\varepsilon_b>0$, $\varepsilon=0$). 
Figure~\ref{fig:diagram}(a) compares conformations at two imposed end-to-end distances: $r_\mathrm{ref}$ for the reference state and $r_\mathrm{str}\,(>r_\mathrm{ref})$ for a stretched state.
Stretching the chain reduces the number of local bends (indicated by pink-filled circles), thereby decreasing the average bend count. 
Consequently, the bending contribution decreases with increasing $r$, resulting in a negative energetic curvature $k_U<0$, as illustrated in Fig.~\ref{fig:diagram}(c). 
In contrast, the corresponding reduction in conformational possibilities lowers the entropy with increasing $r$, so that the entropic curvature remains positive, $k_S>0$, as shown in Fig.~\ref{fig:diagram}(b). 
This competition between energetic and entropic contributions is consistent with the stiffness decomposition in Fig.~\ref{fig:landscape} and complements the soft-repulsion mechanism described in Ref.~\cite{ShiraiSakumichi2025}.

\begin{figure}[t]
\centering
\includegraphics[width=0.99\linewidth]{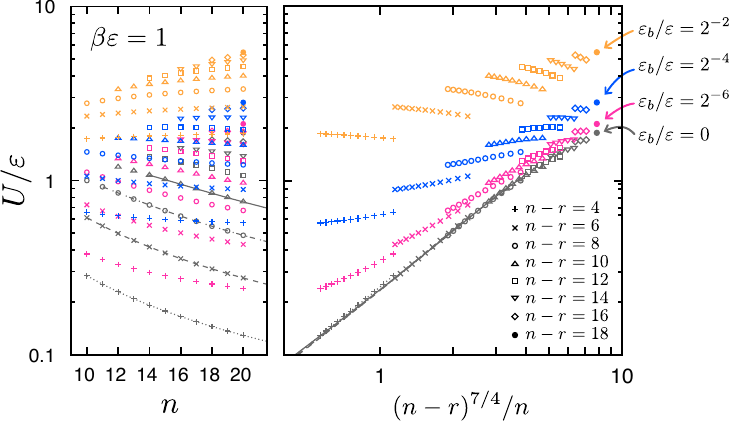}
\caption{
Breakdown of the universal $7/4$-scaling of the internal energy due to bending stiffness.
In the left panel, $U/\varepsilon$ is plotted against $n$ at $\beta\varepsilon=1$ for chains with $n=10,11,\ldots,20$ and slacks $n-r=4,6,\ldots,18$; symbol shape represents $n-r$, and color indicates the ratio $\varepsilon_b/\varepsilon$ of the bending penalty to the self-repulsion strength.
In the right panel, the same data are plotted against $(n-r)^{7/4}/n$, showing how the scaling with slack is affected by bending.
Gray symbols and curves correspond to the DJ model without bending, $\varepsilon_b/\varepsilon=0$, and reproduce the universal $(n-r)^{7/4}/n$ collapse reported in Fig.~5(b) of Ref.~\cite{ShiraiSakumichi2025}.
As $\varepsilon_b/\varepsilon$ increases to $2^{-6}$ (pink), $2^{-4}$ (blue), and $2^{-2}$ (orange), the colored data systematically deviate from the gray master curve, demonstrating that bending stiffness progressively disrupts the universal scaling characteristic of pure self-repulsion.
}
\label{fig:U_scaling}
\end{figure}

While the sign decomposition in Fig.~\ref{fig:landscape} demonstrates that both mechanisms produce negative energetic elasticity, it does not indicate which mechanism predominates in a given system.
To distinguish the microscopic origins of negative energetic elasticity, we examine how the bending-energy contribution systematically modifies the universal scaling behavior of the internal energy observed in the pure DJ model.
In our previous analysis of the DJ model without bending~\cite{ShiraiSakumichi2025}, we showed that $U/\varepsilon$ obeys a single-parameter scaling that collapses onto a universal master curve when expressed as a function of $(n-r)^{7/4}/n$ across a wide range of chain lengths and interaction strengths. 
The $7/4$ exponent, which originates from the underlying self-repulsive single-chain behavior, was found to be robust along the random-walk--self-avoiding walk crossover, and therefore serves as a distinctive signature of negative energetic elasticity driven solely by self-repulsion.
Figure~\ref{fig:U_scaling} extends this scaling analysis to the present model, incorporating bending stiffness. 
For $\varepsilon_b/\varepsilon=0$, the gray symbols reproduce the DJ limit: at fixed $\beta\varepsilon=1$, data for chains with $n=10,11,\ldots,20$ and slack $n-r=4,6,\ldots,18$ collapse onto the same $(n-r)^{7/4}/n$ master curve previously reported in Ref.~\cite{ShiraiSakumichi2025}. 
Introducing finite bending stiffness ($\varepsilon_b/\varepsilon=2^{-6}, 2^{-4}$, and $2^{-2}$) causes the colored data points to deviate systematically from the master curve, with deviations increasing for larger $\varepsilon_b/\varepsilon$ values.
These results indicate that the bending-energy term progressively disrupts the simple $7/4$ scaling that emerges from pure self-repulsion.

This breakdown of universal scaling provides a practical means to distinguish between the two microscopic mechanisms of negative energetic elasticity. 
If experimental or numerical data for $U$ at a fixed end-to-end distance collapse onto a single $(n-r)^{7/4}/n$ curve, the behavior is consistent with solvent-induced self-repulsion as the dominant mechanism. 
In contrast, systematic deviations from this collapse---such as those illustrated in Fig.~\ref{fig:U_scaling} with increasing $\varepsilon_b/\varepsilon$---indicate an increasing contribution from bending stiffness.
Thus, the internal-energy scaling not only unifies the self-repulsive and bending mechanisms at the level of sign but also provides a diagnostic tool to identify which mechanism governs negative energetic elasticity in a given system.

\section{Conclusion}
\label{sec:conclusion}
By exact enumeration of a lattice polymer incorporating DJ self-repulsion and a local bending-energy cost, we have shown that the energetic contribution to the stiffness is negative across the explored parameter space, while the entropic contribution remains positive. 
This finding unifies two previously independent mechanisms of negative energetic elasticity---solvent-induced self-repulsion and bending stiffness---demonstrating that either mechanism alone, as well as their combination, produces the same sign of energetic curvature. 
The Potts/Ising-chain mapping clarifies the bending-only limit, and the stiffness landscape presented in Fig.~\ref{fig:landscape} illustrates how energetic and entropic contributions jointly determine the overall mechanical response.

Beyond the sign of the energetic contribution, our analysis of the internal-energy scaling provides a more sensitive probe of the underlying microscopic mechanism. 
In the absence of the bending-energy term, the internal energy in the DJ limit exhibits a robust $(n-r)^{7/4}/n$ scaling collapse, reflecting the characteristic soft-repulsive single-chain behavior reported previously~\cite{ShiraiSakumichi2025}. 
As shown in Fig.~\ref{fig:U_scaling}, the introduction of bending stiffness progressively disrupts this universal $7/4$ scaling, even though the negative energetic elasticity itself remains intact. 
We therefore propose internal-energy scaling under a fixed end-to-end constraint as a practical diagnostic: successful collapse onto the $(n-r)^{7/4}/n$ master curve indicates that self-repulsion is the dominant source of negative energetic elasticity, whereas systematic deviations induced by bending reflect a significant contribution from bending stiffness. 
In this way, the present model not only unifies the origins of negative energetic elasticity but also provides a practical framework to distinguish the relative roles of self-repulsion and bending stiffness.

Although our analysis is based on exact enumeration of relatively short lattice chains, the qualitative features we identify---namely, the robust negative sign of the energetic stiffness and the sensitivity of internal-energy scaling to bending---are expected to persist in longer chains and more realistic polymer models. 
Future simulations and experiments on gels in which both solvent-induced repulsion and bending stiffness can be tuned would provide a direct test of the proposed diagnostic.

\begin{acknowledgments}
We thank Naoyuki Sakumichi for valuable discussions.
This study was supported by JSPS KAKENHI Grant Nos.~JP25K17313 and JP25K00966.
\end{acknowledgments}

\section*{Data Availability}
The data that support the findings of this article are openly available~\cite{DAS}.

\end{document}